%% file: sample631.tex
\documentclass[twocolumn,]{aastex631}

\usepackage{savesym}
\savesymbol{tablenum}

\usepackage{amsmath}
\usepackage{siunitx}

\usepackage[T1]{fontenc}
\newcommand\lat{{\it Fermi}-LAT}
\newcommand\chandra{{\it Chandra}}
\newcommand\xmm{{\it XMM}-Newton}
\newcommand\xrt{{\it Swift}-XRT}

\DeclareSIUnit\gauss{G}
\DeclareSIUnit\parsec{pc}
\DeclareSIUnit\mas{mas}
\DeclareSIUnit\sr{sr}
\DeclareSIUnit\yr{yr}
\DeclareSIUnit\erg{erg}
\DeclareSIUnit\yr{yr}
\DeclareSIUnit\yrs{yrs}
\DeclareSIUnit\mas{mas}
\DeclareSIUnit\ph{ph}
\DeclareSIUnit\Jy{Jy}

\begin{document}

\title{\lat{} detection of the low-luminosity radio galaxy NGC~4278 during the LHAASO campaign}

\correspondingauthor{Ettore Bronzini}
\email{ettore.bronzini@inaf.it}

\author[0000-0002-0786-7307]{Ettore Bronzini}
\affiliation{Dipartimento di Fisica e Astronomia "Augusto Righi", Università di Bologna, via P. Gobetti 93/2, 40129 Bologna, Italy}
\affiliation{INAF, Astrophysics and Space Science Observatory Bologna, via P. Gobetti 93/3, 40129 Bologna, Italy}

\author[0000-0003-1848-6013]{Paola Grandi}
\affiliation{INAF, Astrophysics and Space Science Observatory Bologna, via P. Gobetti 101, 40129 Bologna, Italy}

\author[0000-0002-5201-010X]{Eleonora Torresi}
\affiliation{INAF, Astrophysics and Space Science Observatory Bologna, via P. Gobetti 101, 40129 Bologna, Italy}

\author[0000-0002-3308-324X]{Sara Buson}
\affiliation{Julius-Maximilians-Universit\"at W\"urzburg, Fakultät f\"ur Physik und Astronomie, Emil-Fischer-Str. 31, D-97074 W\"urzburg, Germany}
\affiliation{DESY, D-15738 Zeuthen, Germany}



\begin{abstract}
We present a study of the high-energy properties of the compact symmetric object NGC~4278, recently associated with a TeV source by the Large High Altitude Air Shower Observatory (LHAASO) collaboration. 
We conducted a dedicated analysis of a \lat{} region around NGC~4278, limited to the LHAASO campaign conducted from March 2021 to October 2022. A statistically significant emission ($\mathrm{TS} = 29$) was revealed, spatially consistent with the radio position of NGC~4278 and the LHAASO source. The \lat{} source is detected above $8\, \si{\giga \electronvolt}$, exhibiting a hard spectrum ($\Gamma=1.3\pm0.3$) and a $\gamma$-ray luminosity of $\mathcal{L}_{>100 \, \si{\mega \electronvolt}} \simeq 4\times 10^{41} \, \si{\erg \, \s^{-1}}$.
A serendipitous \xrt{} observation of NGC~4278 during the TeV campaign reveals the source in a high-state, with a flux $\mathcal{F}_{0.5-8\, \si{\kilo \electronvolt}} = 5_{-2}^{+3} \times 10^{-12} \, \si{\erg \, \second^{-1} \centi \meter^{-2}}$, compatible to the highest luminosity level observed in previous \chandra{} pointings.
The high-energy spectral energy distribution of the source and the intense flux variation observed in the X-ray band support a jet origin for the observed radiation.
We suggest that the significant enhancement of the high-energy flux observed during the LHAASO campaign is due to a transient, highly energetic perturbation in the jet.
The detection of NGC~4278 at both high- and very high-energies opens new frontiers in studying particle acceleration processes. It reveals that even compact, low-power radio galaxies, not just bright blazars, can exceed the sensitivity thresholds of GeV and TeV instruments, becoming promising targets for the upcoming Cherenkov Telescope Array Observatory (CTAO).

\end{abstract}



\input{Introduction}
\input{NGC4278}
\input{Fermi}

\input{X-ray}

\input{Discussion}


\begin{acknowledgments}
The authors thank B. Rani, F. Longo, and M. Kerr for the useful suggestions that improved the clarity of this manuscript. The authors also thank L. Bruno, P. Da Vela, R. De Menezes, G. Migliori, C. Pallanca, and V. Sguera for intensive and constructive discussions. E.B. acknowledges financial support from the European Union – Next Generation EU” RFF M4C2 under the project IR0000012 - CTA+ (CUP C53C22000430006), announcement N.3264 on 28/12/2021: “Rafforzamento e creazione di IR nell’ambito del Piano Nazionale di Ripresa e Resilienza (PNRR)”.
This work was supported by the European Research Council, ERC Starting grant MessMapp, S.B. Principal Investigator, under contract no. 949555, and by the German Science Foundation DFG, research grant 'Relativistic Jets in Active Galaxies' (FOR 5195, grant No. 443220636). The \textit{Fermi}-LAT Collaboration acknowledges support for LAT development, operation and data analysis from NASA and DOE (United States), CEA/Irfu and IN2P3/CNRS (France), ASI and INFN (Italy), MEXT, KEK, and JAXA (Japan), and the K.A.~Wallenberg Foundation, the Swedish Research Council and the National Space Board (Sweden). Science analysis support in the operations phase from INAF (Italy) and CNES (France) is also gratefully acknowledged. This work performed in part under DOE Contract DE-AC02-76SF00515.

\end{acknowledgments}

%







\bibliography{sample631}{}
\bibliographystyle{aasjournal}



\end{document}

%% file: Introduction.tex
\section{Introduction}
\label{sec: introduction}

In the last three decades, very long baseline interferometry (VLBI) imaging surveys \citep[e.g.,][]{Wilkinson1994,Peck2000} have revealed the presence of compact radio sources hosted at the center of many massive galaxies. Among these small, barely resolved objects, compact symmetric objects (CSOs) are characterized by radio emission on both sides of an active galactic nucleus (AGN), linear sizes less than one kpc \citep{Wilkinson1994}, low variability, low apparent speeds along the jets \citep{Kiehlmann2024}, and kinematic ages smaller than a few thousand years \citep{Orienti2016}. 


In the radio band, CSOs are characterized by convex synchrotron spectra, peaking at $\sim 1\, \si{\giga \hertz}$ \citep{ODea1997,Orienti2016,ODea2021}. This spectral characteristic is interpreted in terms of absorption mechanisms, with the emission being optically thin (thick) above (below) the peak frequency \citep[see][for further details]{Snellen2000,Orienti2008,Tingay2015,Callingham2015}.

The nature and compactness of CSOs is still a matter of debate. Among the scenarios proposed to explain their peculiar properties there are: (i) the radio activity has been triggered recently \citep[$\lesssim 100\,$yrs,][]{Owsianik1998a,Owsianik1998b,Tschager2000}; (ii) they might be confined by an extraordinarily dense interstellar medium (ISM) of the host galaxies \citep[e.g.,][]{vanBreugel1984,ODea1991,Dicken2012}; (iii) they might be living an intermittent/transient phase of radio activity, due to instabilities in the accretion disk \citep{Czerny2009}, to accretion of a limited amount of material \citep[e.g., a star,][]{Readhead1994}, or to the disruption of the jet \citep[e.g.,][]{DeYoung1991,Sutherland2007, Wagner2011, Bicknell2018,Mukherjee2018a}. In case of recurrent activity, drops in surface brightness profiles and diffuse radio relics are expected on scales larger than the CSO size \citep[e.g.,][]{Baum1990,Taylor1998,Stanghellini2003,Marecki2003,Maness2004,Stanghellini2005}. The central engine itself is old \citep[$\gtrsim 10^4 \, \si{\yrs}$,][and references therein]{ODea2021}, but the components related to the CSO morphologies, hence the radio source, are young \citep[$\lesssim$ hundreds of years, e.g.,][]{Snellen2000}. 


Studies of CSOs at high energies are fundamental in tracing the most energetic processes in action in these sources. Few theoretical models predict $\gamma$-ray emission from CSOs from GeV up to TeV energies \citep{Stawarz2008,Ito2011,Kino2011,Kino2013,Migliori2014}. At GeV energies, only three CSOs have been detected so far and included in the latest {\it Fermi} Large Area Telescope (\lat{}) Fourth Source Catalog - Data Release 4 \citep[4FGL-DR4,][]{Ballet2023}: PKS 1718-649 \citep{Migliori2016}, NGC 3894 \citep{Principe2020} and TXS 0128+554 \citep{Lister2020}. All sources show no statistically significant variability, and are well described by a simple power law model with a photon index of $\Gamma \gtrsim 2$ and luminosities in the $0.1-100\, \si{\giga \electronvolt}$ band ranging from $10^{42}$ to $10^{47} \, \si{\erg \second^{-1}}$. None of them has been detected above 100 GeV to date.

The association of the Large High Altitude Air Shower Observatory (LHAASO) source 1LHAASO~J1219+2915 with the CSO NGC~4278 \citep{Cao2024a,Cao2024b} represents a remarkable discovery, opening new opportunities to investigate particle acceleration processes in low-luminosity radio galaxies, still confined within their host galaxies. In this Letter, we present a detailed study of NGC~4278 with \lat{} data simultaneous to the first LHAASO campaign (from March 2021 to October 2022), reporting a $\gamma$-ray detection of the source at a significance of $\sim 4.3 \sigma$. Furthermore, we show that, during the same period, the X-ray core of the source was in a high-state, suggesting a significant activity in the nuclear region of NGC~4278. Throughout this work, a flat cosmology with $H_{0} = 67.4 \, \si{\kilo \meter \, \second^{-1} \mega \parsec^{-1}}$ and $\Omega_{\mathrm{m}} = 0.315$ \citep{Planck2020} is adopted.


%% file: NGC4278.tex
\section{NGC~4278}
\label{sec: NGC4278}

NGC~4278 is a nearby \citep[$D_L \simeq 16.4 \, \si{\mega \parsec}$,][]{Tonry2001} early-type galaxy, part of a small group \citep{Garcia1993}, with no evidences of interaction with the lower mass companion galaxy NGC~4283. The nuclear region of NGC~4278 hosts a low-luminosity AGN \citep[$\mathcal{M}_{\mathrm{SMBH}} \simeq 3\times 10^8 \, \mathcal{M}_{\odot}$,][]{Wang2003,Chiaberge2005}, powered by a radiatively inefficient accretion flow \citep{Ho1997,Balmaverde2014}. NGC~4278 is the faintest and most compact source among the known CSOs \citep{Kiehlmann2024,Readhead2024}.

Very Long-Baseline Array (VLBA) observations both at $5 \, \si{\giga \hertz}$ and $8.4 \, \si{\giga \hertz}$ revealed a two-sided radio emission, with a spectral index of $\alpha \gtrsim 0.54$\footnote{Here, spectral indices are defined such that $S \left( \nu \right) \propto \nu^{-a}$, with $S \left( \nu \right)$ being the flux density.}, and emerging from a dominant, flat-spectrum core \citep[$\alpha\sim 0$,][]{Giroletti2005, Tremblay2016}. The overall radio spectrum, that peaks roughly at $1\, \si{\giga \hertz}$ \citep{Giroletti2005, Tremblay2016}, is consistent with synchrotron radiation by accelerated relativistic particles \citep{{Nagar2002}}. The source shows variability in the radio band on yearly timescales, with an enhanced activity at $6 \, \si{\centi \meter}$ around 1985, while a more quiescent activity from 1990 to 2005 \citep{Giroletti2005}.
The edge-darkened FR-I like morphology, with a radio linear size of $\sim 3 \, \si{\parsec}$, suggests type I CSO characteristics \citep[see][for further details]{Tremblay2016,Readhead2024}. The estimated jet-power of the source is modest \citep[$\mathcal{L}_{\mathrm{jet}} \sim 10^{42} \, \si{\erg \, \second^{-1}}$, see][hereafter P12, for details]{Pellegrini2012}. The S-shaped symmetry observed, with the northern jet approaching to us at a mildly relativistic speed ($\beta=v/c \sim 0.75$) and seen at an inclination angle between 2$\si{\degree}$ and 4$\si{\degree}$ \citep{Giroletti2005}, is peculiar, even if not unique for CSOs \citep[e.g.,][]{Baum1990,Readhead1993,Taylor2009}. The distorted radio morphology might be due to beaming effects or, alternatively, to the jet interaction with the surrounding medium \citep{Giroletti2005}. 

From infrared to UV, the overall emission of NGC~4278 is dominated by the host galaxy and dust emission. Its stellar population is old ($> 10\, \si{\giga \yr}$) with no sign of ongoing star formation both in near-UV and mid-IR \citep{Shapiro2010,Kuntschner2010}.
The nuclear emission is unresolved in the optical band \citep{Capetti2000}. \cite{Younes2010} detected optical nuclear variability up to a factor of 4 between December 2006 and January 2007 using the \textit{Hubble} Space Telescope (HST). In the UV band, the nucleus is barely resolved and rapidly variable \citep{Cardullo2009}, as commonly observed in most low-ionization nuclear emission-line region (LINER) galaxies \citep{Maoz2005}: an enhanced flux by a factor of 1.6 was observed between June 1994 and January 1995 \citep{Cardullo2009}.

NGC~4278 has been extensively studied in the X-ray band with \chandra{} and \xmm{} \citep{Ho2001, Terashima2003,Gonzalez2009,Fabbiano2010,Younes2010,Hernandez2013}. \chandra{} images of NGC~4278 are dominated by the nuclear point-like emission, whose best-fit model is a superposition of a power law spectrum ($\Gamma=2.31\pm0.20$) and a thermal component ($kT = \left(0.75 \pm 0.05\right) \, \si{\kilo \electronvolt}$), with a low level of obscuration ($N_\mathrm{H} = \left(4.2 \pm 3.1\right) \times 10^{20} \, \si{\centi \meter^{-2}}$, P12). The nuclear emission shows a variability in \chandra{} images, with a luminosity of the power law component in the range $\mathcal{L}_{0.5-8 \, \si{\kilo \electronvolt}} \simeq 0.3-6 \times 10^{40} \, \si{\erg\, \second^{-1}} $.

At $\gamma$-ray energies, NGC~4278 is not reported in the latest 4FGL-DR4 catalog \citep{Ballet2023}, nor in the previous ones. 
It was revealed during a one-month flare from March 5 to April 5, 2009, and is listed in the First Catalog of Long-term Transient Sources \citep[1FLT,][]{1FLT}, based on the first 10 years of \lat{} data. On that occasion, the source was detected with a significance of about $5 \sigma$, a photon flux of $\mathcal{F}_{0.1-300 \, \si{\giga \electronvolt}} = \left(9.8\pm 3.1\right) \times 10^{-8} \, \si{\ph \, \second^{-1} \, \centi \meter^{-2}}$, and a photon index $\Gamma = 3.3 \pm 0.4$. However, the probability that this event is a false positive detection is non-negligible \citep[$P\sim34\%$,][]{1FLT}.

NGC~4278 has been suggested as a possible counterpart of 1LHAASO~J1219+2915, a very high-energy (VHE) $\gamma$-ray source detected by the Large High Altitude Air Shower Observatory (LHAASO) Water Cherenkov Detector Array (WCDA) and listed in the first LHAASO catalog \citep{Cao2024a}. This catalog was compiled using 508 days of data collected by the WCDA from March 2021 to September 2022, and 933 days of data recorded by the KM2A from January 2020 to September 2022. However, we note that 1LHAASO~J1219+2915 was not detected by KM2A, therefore, in the following, we will refer only to WCDA data. More recently, an extended study up to the end of October 2023 \citep{Cao2024b}, leading to an effective live-time of about 891 days, was performed. A point-like emission located at $\sim 0.03\si{\degree}$ ($\mathrm{R.A.}=185.05\si{\degree}\pm 0.04\si{\degree}$, $\mathrm{Dec.}=29.25\si{\degree} \pm \si{ 0.03\degree}$) from the radio position of NGC~4278 is detected at an overall significance of $6.1 \sigma$. Interestingly, the source showed hints of variability ($p\mathrm{-value}=2.6\times 10^{-3}$), with an enhanced TeV activity, from the middle of August 2021, to the middle of January 2022, of about 7 times the low-state \citep{Cao2024b}. During the active phase, the source was detected at $7.8\sigma$ and the spectrum is well described in the $1-15 \, \si{\tera \electronvolt}$ energy band by a single power law model, with a photon index of $\Gamma = 2.56 \pm 0.14$, and a flux of $\mathcal{F}_{1-10 \, \si{\tera\electronvolt}} = \left(7.0 \pm 1.1_{{\rm stat.}} \pm 0.4_{{\rm syst.}} \right) \times 10^{-13}\, \si{\ph \, \centi \meter^{-2} \s^{-1}}$ \citep{Cao2024b}. 
The proximity of NGC~4278 makes the $\gamma\gamma$ absorption due to the extragalactic background light negligible.


%% file: Fermi.tex
\section{\lat{} Analysis}
\label{sec: fermi}
The discovery of TeV photon emission in a very low luminosity radio galaxy has prompted the immediate question whether its emission may extend also to GeV energies, and be observable by \lat{}. Our aim was twofold: 1) to corroborate the TeV detection; 2) to extend the spectral study to lower $\gamma$-ray energies in order to shed light on the physical processes at work.

Unlike other attempts to search for GeV signals from the source based on 14 years of \lat{} data recently reported in the literature \citep{Wang2024,Lian2024}, our approach was to limit the investigation to the time interval corresponding to the first LHAASO catalog \citep{Cao2024a}.

The \lat{} is a $\gamma$-ray telescope that detects photons by conversion into electron-positron pairs. Its operational energy range extends from $\sim 50 \, \si{\mega \electronvolt}$ up to $\sim 1 \, \si{\tera \electronvolt}$. The \lat{} is equipped with a high-resolution converter tracker (for direct measurement of the incident $\gamma$-rays), a CsI(Tl) crystal calorimeter (for energy measurement), and an anti-coincidence detector to reject the background of charged particles \citep{Atwood2009}.

We inspected the \lat{} region around NGC~4278, integrating data from March 1\textsuperscript{st}, 2021 (MJD=59274) to October 1\textsuperscript{st}, 2022 (MJD=59853). We considered both front- and back-converted events in the $100 \, \si{\mega \electronvolt}-1\, \si{\tera \electronvolt}$ energy range. We selected the  \texttt{SOURCE} class events and the \texttt{P8R3\_SOURCE\_V3} instrument response functions \cite[IRFs,][]{Bruel2018}.

The data analysis was performed by exploiting the open-source python package \textit{Fermipy} v.1.2.0 \citep{Wood2017} with \textit{Fermitools} v.2.2.0. 
Only good time intervals (\texttt{LAT\_CONFIG==1} and \texttt{DATA\_QUAL>0}) were selected, resulting in an exposure time of $\sim1.3 \, \si{\yrs}$. We verified that the Sun and the Moon were at a distance greater than $20\si{\degree}$ from NGC~4278 and therefore do not affect our data.

Taking advantage of the event partition introduced with \textsc{Pass 8} \citep{Atwood2013}, we analyzed separately the four different event classes, each characterized by its own point spread function (PSF). Each PSF indicates the quality of the reconstructed direction of the event, from the best (PSF3) to the worst (PSF0). Following the procedure outlined in the 4FGL-DR3 catalog \citep{Abdollahi2022}, we further divided the data into 6 different energy bands. The number of bins and the optimal zenith angle selection remain consistent across each energy interval, while each event type necessitates a different pixel size depending on the PSF width \citep[refer to Table 1 in][]{Abdollahi2022}. Differently from the 4FGL-DR3 catalog, we additionally split the $30\, \si{\giga \electronvolt}  - 1 \, \si{\tera \electronvolt}$ energy bin into the four PSF components. In total, we dealt with 21 components.

A circular region of interest (RoI) of radius $15\si{\degree}$ centered on the radio position of NGC~4278 \citep[$\mathrm{R.A.}=185.03\si{\degree}$, $\mathrm{Dec.}=29.28\si{\degree}$,][]{Ly2004} was chosen. The model included the most recent template for Galactic interstellar diffuse emission (\texttt{gll\_iem\_v07.fits}), the isotropic background components appropriate for each PSF (\texttt{iso\_P8R3\_SOURCE\_V3\_PSFx\_v1.txt})\footnote{See \url{https://fermi.gsfc.nasa.gov/ssc/data/access/lat/BackgroundModels.html} for further details.}, and the sources listed in the 4FGL-DR4 catalog \citep{Ballet2023}. The energy dispersion correction (\texttt{edisp\_bins=-2}) is enabled for all sources except the isotropic component. The test statistic (TS) was used to estimate the significance of a source\footnote{The TS corresponds to the logarithmic ratio of the likelihood of a model with the source being at a given position in a grid ($\mathcal{L}_{\mathrm{src}}$) to the likelihood of the model without the source ($\mathcal{L}_{\mathrm{null}}$), $\mathrm{TS}=2\log \left(\mathcal{L}_{\mathrm{src}}/\mathcal{L}_{\mathrm{null}} \right)$ \citep{Mattox1996}. }. Preliminarily, a fast optimization of the entire RoI was performed, deriving the best-fit model of the RoI. As NGC 4278 is not included in the 4FGL-DR4 catalog, we included a point-source at the radio coordinate position of the target, and parametrized it with a power law model with spectral index initially fixed to 2. Then, the parameters of both diffuse components along with the spectral parameters of sources within a $5\si{\degree}$ radius surrounding the source of interest were set to vary during the fit. Only the normalizations were left free to vary for sources at a distance between $5\si{\degree}$ and $10\si{\degree}$ from the center of the RoI. The spectral parameters of sources at larger distances, as well as the faintest ones ($\mathrm{TS}<4$), were kept fixed. A binned likelihood fit, minimized using \texttt{Minuit}, was performed for all the 21 components separately, and then summed to obtain the total log-likelihood. After performing the fit, we searched for the best position of the source detected at the center of the RoI. The peak of the emission was found at $\mathrm{R.A.}=184.99\si{\degree} \pm 0.02$ and $\mathrm{Dec.}=29.28 \si{\degree} \pm 0.03$, approximately $1\sigma$ away from the radio position of NGC~4278. The significance of the source detection results to be $\mathrm{TS}=29$, corresponding to $\sim 4.3 \sigma$ for 4 degrees of freedom, i.e., the two spectral parameters plus the coordinates.

NGC~4278 exhibits a hard photon index ($\Gamma = 1.3 \pm 0.3$) and an energy flux $\mathcal{F}_{>100\, \si{\mega \electronvolt}} = \left(1.2\pm 0.9\right) \times 10^{-11} \, \si{\erg \, \second^{-1} \, \si{\centi\meter^{-2}}}$, corresponding to an isotropic luminosity of $\mathcal{L}_{>100 \, \si{\mega\electronvolt}} = \left(4\pm 3 \right) \times10^{41} \, \si{\erg \, \second^{-1} }$.

\begin{figure}
    \centering
    \includegraphics[width=0.47\textwidth]{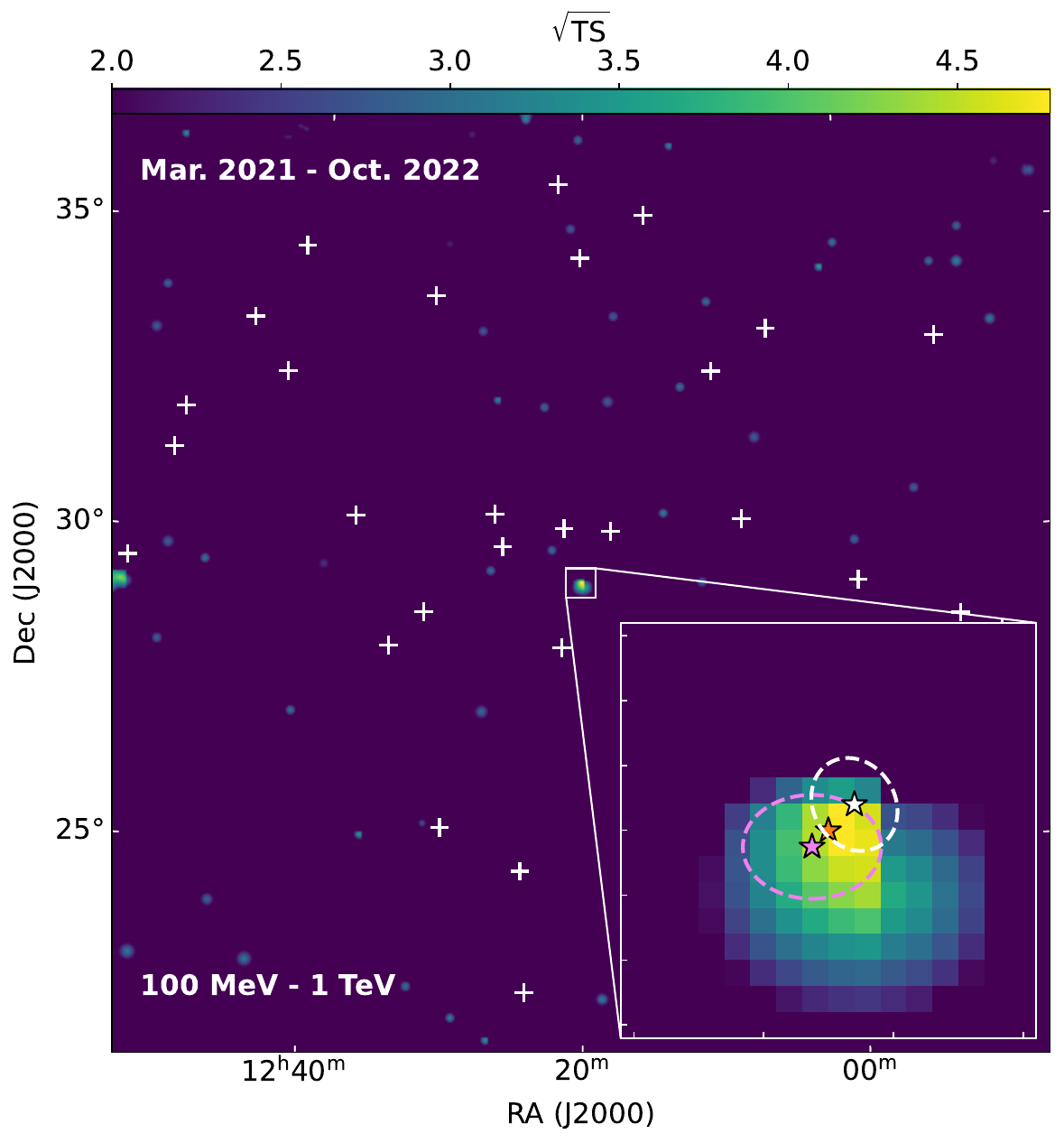}
    \caption{TS map of NGC~4278. The position of the 4FGL-DR4 sources are shown as white crosses. In the zoomed-in map, the orange star marks the radio position of NGC~4278, while the violet and white ones are the LHAASO and \lat{} best-fit positions, respectively. The LHAASO and \lat{} 95\% localization uncertainty regions are shown as dashed circular regions using the same color code.}
    \label{fig: ts_map}
\end{figure}

The resulting localization is shown in TS map in Figure \ref{fig: ts_map}, that was generated using a power law spectral model with a photon index of 2. The orange star marks the radio position of NGC~4278, while the violet and white ones are the LHAASO and \lat{} best-fit positions, respectively. The LHAASO and \lat{} 95\% localization uncertainty regions are shown as dashed circular regions using the same color code.

To consolidate our analysis, we used the \texttt{gtsrcprob} tool to estimate the probability of each event to be associated with NGC~4278. We observed that the detection is driven by three high-energy events, all front-converted, with a probability of more than 90\% of being associated with NGC~4278 (see column 6 in Table \ref{tab: photons}). Specifically, we detected one high-energy photon at 8 GeV and two very high-energy photons at 96 and 103 GeV, both having a probability exceeding 99\%.

\input{photons}

We also checked the quality for the reconstructed events. All the three events tracks in the detector are well tracked and the reconstruction of the incoming direction is precise: the events at 8 and 103 GeV belong to PSF2, while the event at 96 GeV belongs to PSF3. The reconstructed arrival direction of each event is listed in columns 3 (R.A.) and 4 (Dec.) in Table \ref{tab: photons}. 
Furthermore, we inspected the event class associated to each of the three photons. All the events pass the \texttt{CLEAN} quality cut. The two VHE photons even belong to the \texttt{SOURCEVETO} class (see column 5 in Table \ref{tab: photons}). 
The associated classes exclude with very high probability that the detected photons are due to the residual background \citep[see][for further details]{Bruel2018}.

As a final test, the analysis was repeated selecting only events belonging to the purer \texttt{CLEAN} and \texttt{SOURCEVETO} classes, respectively.
The results are completely consistent with the previous ones.

%% file: photons.tex
\begin{table}
    \footnotesize
    \centering
    \begin{tabular}{rcccccc}
    \hline
    \hline
        Energy & MJD & $\alpha_{\mathrm{J200}}$ & $\delta_{\mathrm{J200}}$ & \texttt{evclass} & \texttt{gtsrcprob}\\
        $\left[\si{\giga \electronvolt}\right]$ & & $\left[\si{\deg}\right]$ & $\left[\si{\deg}\right]$ & & $\left[ \%\right]$\\
        (1) & (2) & (3) & (4) & (5) & (6)\\
        \hline
        8 & 59561 & 184.98 & 29.36 & \texttt{CLEAN} & 93.8 (3.8, 2.1) \\
        96 & 59742 & 185.03 & 29.28 & \texttt{SOURCEVETO} & 99.9 (<1, <1)\\
        103 & 59312 & 184.95 & 29.20 & \texttt{SOURCEVETO} & 99.2 (<1, <1) \\
    \hline
    \end{tabular}
    \caption{Properties of photons associated to NGC~4278 with a probability higher than 90\%. \\
    \textit{Notes.} Column (1) reconstructed energy, (2) arrival time in modified Julian day (MJD), (3-4) reconstructed right ascension and declination, (5) event class of the reconstructed event (further details in the text), (6) association probability using the \texttt{gtsrcprob} tool for NGC~4278 and for the backgrounds in parentheses (isotropic and galactic, respectively).}
    \label{tab: photons}
\end{table}

%% file: X-ray.tex
\section{X-ray observations during the LHAASO campaign}

NGC~4278 has been extensively studied in the X-ray band. Thanks to its proximity, \chandra{} could resolve both the nuclear and galactic emissions \citep{Brassington2009,Fabbiano2010,Younes2010}. As shown by P12, the nucleus of this source is generally dominant and highly variable, with flux variations by a factor of $\sim$18 from 2004 to 2010 in the $0.5-8 \, \si{\kilo \electronvolt}$ band.

No other pointings of the source are available in the public archives. Only \xrt{} serendipitously observed NGC~4278 for about $1 \si{\kilo \second}$, on November 28, 2021, within the first LHAASO campaign. 
The online tool for \xrt{} \citep[see][for details]{Evans2009} was used to analyse these data.
{\it Swift}-XRT has a limited angular resolution ($\sim18\si{\arcsecond}$ at $1.5 \, \si{\kilo \electronvolt}$) with respect to \chandra-ACIS ($\sim0.5\si{\arcsecond}$ at $1.5 \, \si{\kilo \electronvolt}$), that prevented us from performing a spatially resolved analysis. However, taking advantage of the detailed \chandra{} analysis by P12, we estimated the contribution to the X-ray spectrum from all the components (central AGN, LMXBs, thermal gas, etc.) falling within an area with a radius of $35.4 \si{\arcsec}$ centered on the source, corresponding to the \xrt{} extraction region. The total number of counts was 27, so we adopted the C-statistics \citep{Cash1979} and rebinned the data to have one count per energy bin. We allowed only the power law parameters to vary and kept the spectral parameters and fluxes of the other components constant.  We did not consider the contribution of unresolved AB+CV stars because it is negligible (see Figure 7 in P12). The intrinsic absorption was fixed to $N_{\mathrm{H}} = 4 \times 10^{20}  \si{\centi \meter^{-2}}$ (P12).

Adopting the Levenberg-Marquardt minimization technique, the best fit resulted in a nuclear spectral slope of $\Gamma = 1.4_{-0.6}^{+0.6}$, and an unabsorbed flux of $\mathcal{F}_{0.5-8\, \si{\kilo \electronvolt}} = 5_{-2}^{+3} \times 10^{-12} \, \si{\erg \, \second^{-1} \centi \meter^{-2}}$ corresponding to a luminosity $\mathcal{L}_{0.5-8\, \si{\kilo \electronvolt}} = 1.6_{-0.6}^{+0.9} \times 10^{41} \, \si{\erg \, \second^{-1}}$. 



The \xrt{} observation shows that the source was in a high-state, with a flux comparable to the highest nuclear levels recorded by \chandra{} (P12).

%% file: Discussion.tex
\section{Discussion and conslusions}

In this Letter, we report a $4.3\sigma$ \lat{} detection of the low-luminosity radio galaxy NGC~4278, recently detected in the TeV band by LHAASO. We notice that NGC~4278 was included in the 1FLT catalog \citep{1FLT} with a high probability ($P\sim 34\%$) of being a false positive. Therefore, we consider our result as the first solid detection of NGC~4278 in the GeV band. 

It is likely that NGC~4278 periodically ejects relativistic plasma in the form of blobs \citep{Giroletti2005}. This hypothesis finds support in its radio spectrum, which shows a flattening below the peak frequency $1\,\si{\giga \hertz}$ \citep{Tremblay2016}.
The low-frequency radiation could be related to a radio relic, which remains after the old jets have dissipated their kinetic energy in the interstellar medium.
The total flux of NGC~4278 at $144 \, \si{\mega \hertz}$ recently reported by the LOw-Frequency ARray \citep[LOFAR,][]{vanHaarlem2013} Two-metre Sky Survey \citep[LoTSS,][]{Shimwell2017} is consistent with the flattening of the radio emission. Moreover, the LOFAR high-resolution ($6\si{\arcsecond}$) image was fitted with more than a single Gaussian component, hinting potential extension of the source, possibly related to plasma located up to scales larger than $\sim 500\,\si{\parsec}$.
\chandra{} also provided further evidence. P12 noted that the gas in the innermost region (within $\si{150\, \parsec}$) is hotter than the ISM of the host galaxy. As suggested by the authors, this additional heating of the inner region may be due to the expansion of the jets in the ISM.
The recent detection of NGC~4278 at TeV energy could represent a renewed activity of this source.


The \lat{} detection of NGC~4278 and the high \xrt{} X-ray flux during the LHAASO campaign provide further hints on the nuclear activity of this source. On the other hand, the large flux variations observed by \chandra{} suggest an origin within the jet of the observed high-energy emission, possible related to some form of jet perturbation (e.g., the injection of a new component in the ejection flow or an internal shock). 


\begin{figure}
    \centering
    \includegraphics[width=0.47\textwidth]{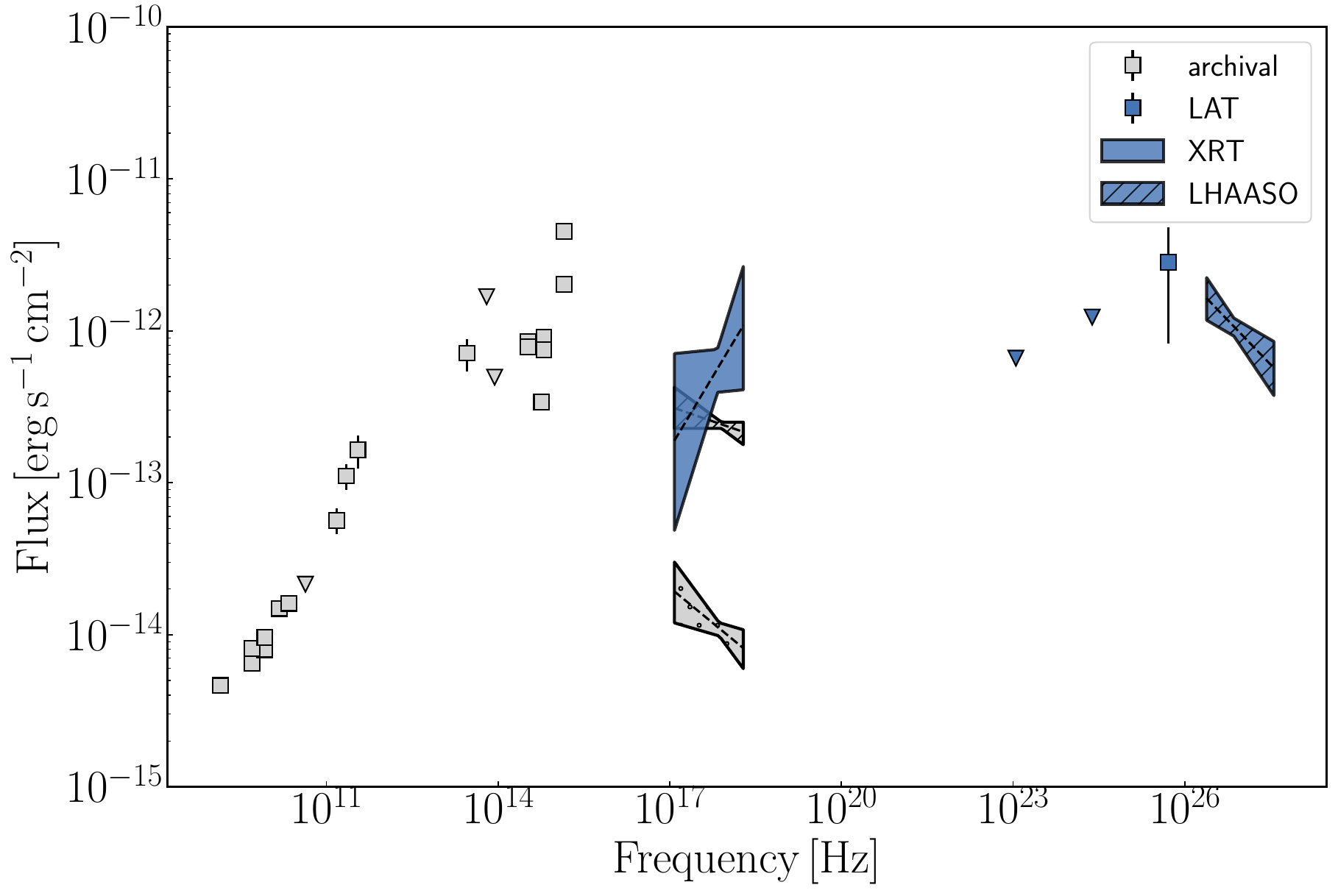}
    \caption{Multiwavelength and multi-epoch SED of nuclear emission in NGC~4278.\\
    \emph{Notes.} Archival data are shown in grey and data are collected from \cite{Anton2004,Giroletti2005,Cardullo2009,Younes2010,Pellegrini2012,Isbell2021}. {\it Swift}-XRT, \lat{}, and LHAASO simultaneous data are shown in blue and are from \cite{Cao2024b} and this work.
    }
    
    \label{fig: SED}
\end{figure}

In Figure \ref{fig: SED}, we show the multi-wavelength and multi-epoch spectral energy distribution (SED) of the nuclear emission in NGC~4278.
It is notable how the \lat{} data seem to naturally extrapolate the TeV spectrum, showing a peak at $\sim 10^{26} \, \si{\hertz}$. This spectral feature resembles the high-energy bump often associated with inverse Compton radiation from jets in a leptonic scenario.
However, the \xrt{} X-ray emission does not connect smoothly with the \lat{} data (see Figure \ref{fig: SED}). Its extrapolation to higher energies lies above the 3$\sigma$ frequentist upper limits by \lat{}. This could indicate a synchrotron origin of the keV radiation, probably peaking in the hard X-ray band \citep[$\gtrsim  10^{18} \, \si{\hertz}$,][]{Harris2006,Georganopoulos2016}, and resembling high-synchrotron peaked (HSP) BL Lac SEDs \citep[see, e.g.,][]{Ghisellini2017}. 

Although a synchrotron self-Compton one-zone model is promising for fitting the \xrt{}, \lat{}, and LHAASO data of NGC~4278, it seems to falter when extended below $10^{17} \, \si{\hertz}$ (see Figure \ref{fig: SED}). The lower energy SED appears apart and unrelated to the high-energy emission. Similar conclusions, although based on a slightly different data set of NGC~4278, were reached by \cite{Lian2024} and \cite{Dutta2024}. The low-energy SED, constructed from non-simultaneous data found in the literature, likely represents the average state of NGC~4278's nucleus. Therefore, it is reasonable to hypothesize the higher energy data as evidence of a transient, highly energetic event occurred during the LHAASO campaign.

Unfortunately, high angular resolution radio observations of the source during and after the LHAASO campaign are not currently available in the literature. To support the proposed picture, new radio observations are mandatory to ascertain potential radio structure perturbations.

In conclusion, the first detection of NGC~4278 at GeV and TeV energies paves the way to studying a new population of radio galaxies similar to HBL at high energies.
This discovery demonstrates that even low-power objects can accelerate particles up to TeV energies, opening a new field of investigation for the Cherenkov Telescope Array Observatory (CTAO).